\documentclass[12pt]{article}

\usepackage{times}
\usepackage{graphicx}
\usepackage{color}
\usepackage{ccaption}
\usepackage{verbatim}
\usepackage{mathtools}
\usepackage{versions}
\usepackage{xr}
\usepackage{hyperref}

\usepackage[labelfont=bf]{caption}

\usepackage [autostyle, english = american]{csquotes}
\MakeOuterQuote{"}
\usepackage{setspace}
\newenvironment{sciabstract}{%
\begin{quote} \bf}
{\end{quote}}

\title{Substrate stiffness governs dynamics and self-organization of nascent biofilms}

\author
{Garima Rani$^{1,2}$ and Anupam Sengupta$^{1,3}$\\
\\
\normalsize{$^{1}$Physics of Living Matter Group, Department of Physics and Materials Science, }\\
\normalsize{University of Luxembourg, 162 A, Avenue de la Fa\"{i}encerie, L-1511, Luxembourg}\\
\normalsize{$^{2}$ Department of Biochemical Engineering and
Biotechnology,}\\
\normalsize{Indian Institute of Technology Delhi,  Hauz Khas, Delhi 110016, India}\\
\normalsize{$^{3}$Institute for Advanced Studies, University of Luxembourg,}\\
\normalsize{2, Avenue de l’Université, L-4365, Esch-sur-Alzette, Luxembourg}\\
}

\date{}
\includeversion{v1}
\excludeversion{v2}

\begin{document} 



\maketitle


\begin{sciabstract}
The evolutionary success of bacteria lies in their ability to form complex surface-associated communities in diverse biophysical settings. From hard to compliant surfaces, bacteria encounter and successfully colonise vastly different substrates which are relevant for both environmental and biomedical contexts. However, it remains poorly understood how compliance of soft surfaces -- measured in terms of their elastic deformability -- impacts the dynamics and self-organization of bacterial cells proliferating into colonies. Using experiments and biomechanical modelling, here we zoom into the expansion and self-organization of bacterial cells into sessile colonies on soft substrates. The dynamics and spatio-temporal structures were captured by visualising growing bacterial colonies on nutrient-rich, soft agarose pads, whose elastic modulus was tuned between $\sim 0.3$ kPA to $\sim 100$ kPA by varying the concentration of the agarose in the underlying substrate. Our results show that, at the scale of the colonies, significant differences emerge in the spreading dynamics and colony geometry as the substrate stiffness is altered: softer substrates promote distinct, multilayered colony structures, and as revealed by fractal analysis of the colony boundaries, they exhibit higher boundary roughness. In contrast, colonies growing on harder substrates first grow up to large monolayers, before undergoing the mono-to-multilayer transition (MTMT), showing nearly $300\%$ increase in the overall colony area at MTMT as substrate stiffness increased within the aforementioned range. A simple biomechanical model captures the role of effective drag forces at different scales, acting on the colonies as they spread on substrates with different stiffness: higher drag in soft substrates drive early verticalisation of the colonies, while lower effective drag delays the MTMT, resulting in larger colony areas. Based on the results and biomechanical insights, a comprehensive data-backed numerical model is currently being developed. Our findings highlight the role of surface stiffness in determining the dynamics, structure and self-organization of bacterial cells into an expanding multi-scale colony.    

\end{sciabstract}

\maketitle

\section*{Introduction}

Bacteria thrive in biophysically diverse environments, spanning terrestrial and aquatic ecosystems, as well as extreme environmental conditions. From the human gut to soil, and glaciers to radioactive wastes, bacteria are one of the most prolific colonisers, thanks to their ability to colonise various surfaces, making this a central quest in microbiology \cite{Costerton95,Jin24}. Surface colonisation is crucial for species growth, that allows refuge and protects bacteria from detrimental conditions, predation as well as promoting optimal use of resources for maximising growth. Indeed, surface colonisation for microbial organisms is arguably as important for their evolutionary success as was the transition of humans from a nomadic lifestyle to settlements, leading to our preeminent position in the biosphere today \cite{Worldsbook}. Akin to how settlement location has affected distinct ways of living in humans, it is not unexpected that surface properties have been shown to greatly influence bacterial, and in general, cellular behavior and ecology \cite{Katsikogianni20048, Tuson2013, Ahmed2015,Schamberger2023}.

For bacteria, early stages of colonisation is a critical determinant of the long-term behavior and survival. Surface attachment, followed by the transition from a two dimensional monolayer of cells to a three dimensional multilayer structure \cite{Beloin2008}, often referred to as mono-to-multi-layer transition (MTMT) \cite{You2019}, constitute two key steps of bacterial colonization. These early stages of biofilm formation has been studied in detail over the last decade, particularly from a mechanobiological viewpoint \cite{You2019, Wingreen2018,Dhar2022,Khan2024}. In this context, it has been understood that, depending on the species in consideration \cite{Hug2017,Courtney2017,Spengler2024,Nezio2024}, surface properties and geometry can play a crucial role in determining effectiveness of microbial colonisation. A range of surface properties and their effects on bacterial colonisation dynamics have been studied \cite{Tuson2013, Song2015,yin2021}. For instance, surface roughness at micro and nano scales have been shown to influence bacterial adhesion due to variable contact areas \cite{Mitik2009}. Other such properties include hydrophobicity \cite{Doyle2000} and surface charge \cite{Hayles2025}, which can also influence how bacteria colonise surfaces. Complementing these studies, recent works have indicated the role of surface curvature on bacterial organisation and distribution of biomechanical stresses, particularly during the early biofilm formation. For instance, in both host-microbe settings as well as in colonisation of a passive substrate, bacteria-surface interactions can be mediated by the local curvature of the system \cite{Schamberger2023}. Considering that a major source of infections in humans are due to surface transmissions of bacteria, a clear understanding of how various surface properties promote or inhibit bacterial colonisation will be key to engineering surfaces to tune bacterial self-organisation into colonies, for instance, via superhydrophobic surfaces \cite{Deepu2023}, topological features \cite{Sengupta2020,Shimaya2022}, or by harnessing confinements \cite{Nuno2023}.   

 Stiffness of the underlying substrates is a crucial determinant of bacteria-surface interactions, with multiple studies demonstrating their impact on bacterial adhesion, growth, morphological transitions as well as antibiotic susceptibility \cite{Saha2013,Lichter2008,Song2014,Grant2014, Mu2023} have focused on colony-scale dynamics. At high cell densities, mechanical constraints imposed by the environment can even lead to glass-like behavior in bacterial populations, as recently observed in dense, quasi-two-dimensional \textit{Escherichia coli} colonies \cite{Lama2024}. Large, colony-scale geometries of bacterial biofilms like those formed by the  \textit{Bacillus subtilis} have been long studied, supported by phase diagrams of their structures with respect to the nutrient medium and substrate stiffness \cite{MATSUSHITA1990}. Recent studies have shown that bacterial behavior encapsulated in soft, 3D hydrogel environments depends on both mechanical and chemical cues \cite{{kandemir2018,Bhusari2022,Lewis2022}}. In mature E. coli biofilms, substrate composition and nutrient availability have also been shown to affect internal architecture, including intra-colony channel morphology \cite{Bottura2022}. These findings highlight the importance of environmental factors in shaping biofilm structure at later stages. However, the role of substrate stiffness specifically during the early stages of a biofilm remains largely unexplored. Most existing research focuses on mature colonies comprising thousands of cells, where variables such as nutrient depletion, internal stress regulation, and accumulation of inactive cells introduce considerable complexity and obscure the direct effects of substrate mechanics \cite{Allen_2019,Wittmann2023}.

Here, we combine experiments and a biomechanical model to delineate how substrate stiffness impacts cellular organization and expansion of nascent biofilms. Starting with single founder microcolonies of sessile \textit{Escherichia coli} bacteria on nutrient-rich, compliant agarose substrates. The stiffness of the compliant substrates was varied by tuning the concentration of the low-melting-point (LMP) agarose used in the substrates, thereby modulating their elastic modulii. We probe the effect of substrate stiffness on the expansion of early-stage biofilms, quantifying the spatio-temporal evolution of the colony geometry, at 2D planar, 3D and fractal dimensions. Our results reveal a subtle influence of substrate compliance in nascent biofilms, which amplify over time to produce marked differences in the colony geometry and mono-to-multilayer transition dynamics over longer timescales. Specifically, our results uncover that stiffer substrates promote morphologically isotropic nascent colonies, with a slower transition to the multi-layered structures; whereas soft, compliant substrates drive anisotropic colony formations, combined with faster 2D to 3D transitions. The contrasting observations point to higher effective drag forces on soft substrates, due to deformation-mediated impediment to the colony expansion. However, fractal analysis reveals an interesting dichotomy, with colonies on softer substrates displaying higher boundary roughness, driven by instabilities emerging at the boundary as the colony grows and cells push outward. Biophysical modeling reveals the fundamental role of drag forces in opposing the movement of cells in expanding colonies across scales, and in determining the resulting variations in colony-scale spreading dynamics. Overall, our results uncover the multi-faceted role of compliant surfaces in mediating the cell-substrate interactions, ultimately governing the way cells self-organise in nascent biofilms before developing into complex multiscale structures.


\section*{Results} 
We investigate the growth dynamics and self-organisation of individual cells to nascent biofilms depending on the stiffness of the underlying soft substrates. We grow surface-associated \textit{E. coli} on specially designed substrates with varying levels of LMP agarose concentration (see Fig. \ref{fig:FIG1_MAIN}(A) and Materials and Methods). The agarose concentration determines the mechanical properties, specifically Young's modulus, which we have measured using atomic force microscopy and in reported in Ref. \cite{Rene2024}. In general, differential agarose concentrations can result in differences in properties like diffusivity \cite{Bochert2022} and viscosity \cite{agaroseProps} as well. Here we focus on surface stiffness and its impact on colony evolution in the early stages of its formation.


\subsection*{Substrate stiffness modulates development of nascent biofilms}
\begin{figure}
\centering
\includegraphics[width=\columnwidth] {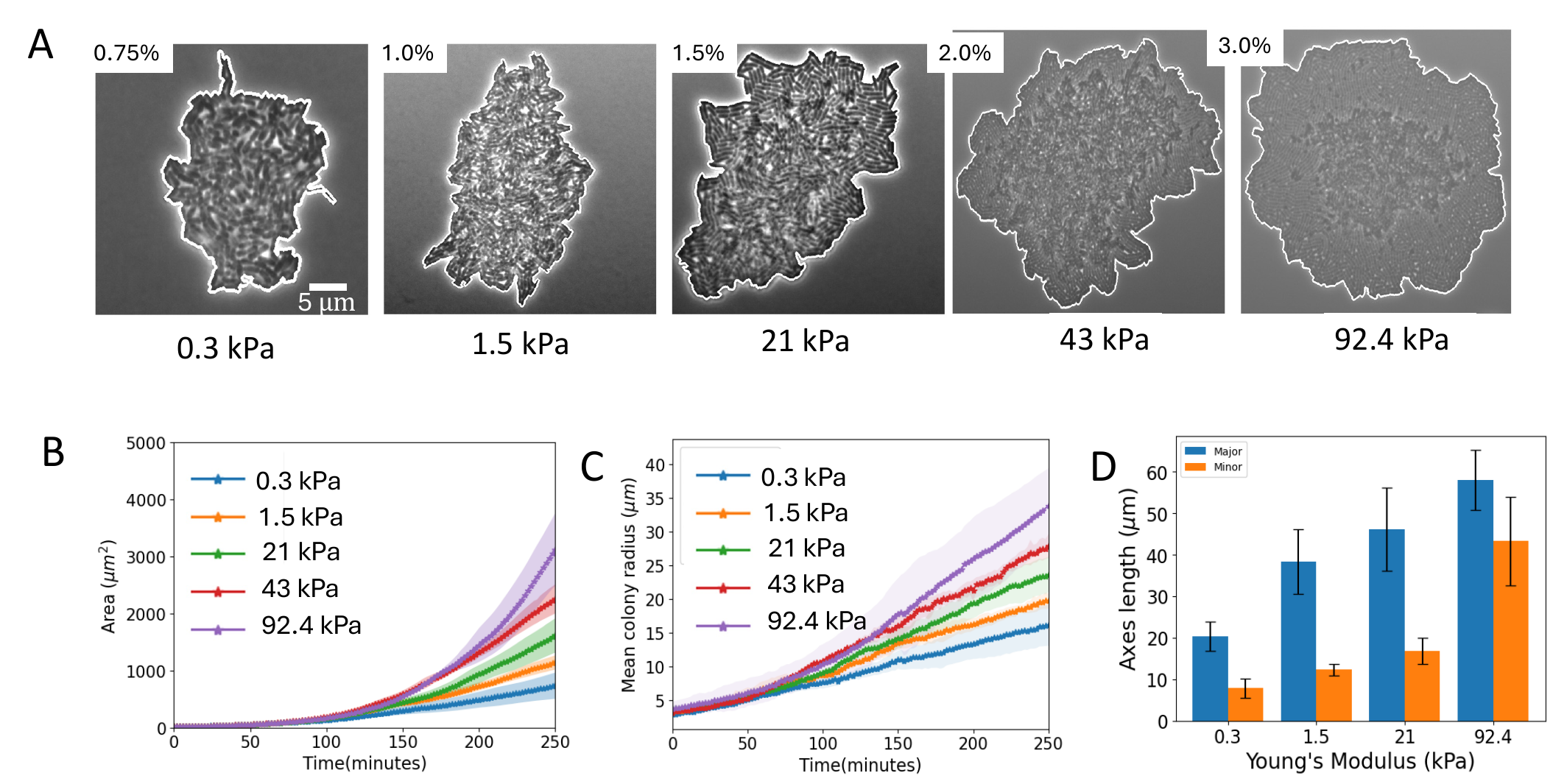}
\caption{\textbf{Morphology of growing \textit{E. coli} colonies on low melting agarose pads, with increasing substrate stiffness.} (A) Snapshots at 6 hours of colony growth on substrates having different concentration of low-melting-point agarose (LMP agarose), as captioned in the images. Here $x\%$ denote $x$ g of LMP agarose in 100 ml of LB media. (B) Colony area is plotted as a function of time. (C) Mean colony radius is plotted as a function of time, (D) The major (blue) and minor (orange) axes of the colonies at MTMT is shown. A significant difference is noted when the Young's modulus of the substrate increases from 21 kPa (low) to 92.4 kPa (high). Data represent at least three biological replicates for each stiffness condition ($n \geq 3$). Statistical difference tested for two sample t-test, p-value $<$ 0.01.}
\label{fig:FIG1_MAIN}
\end{figure}

We first quantify the growth dynamics and colony-scale geometry to understand how substrate stiffness impacts these fundamental paramters. We track the colony growth in terms of relevant size, specifically, the areal growth as well as the mean colony radius (Fig. \ref{fig:FIG1_MAIN}(B) and Fig.\ref{fig:FIG1_MAIN}(C)). We observe a subtle variation wherein initially, across all cases, the colonies showed similar growth dynamics; however, as they grow over time, the spreading dynamics displayed variation, with colonies spreading horizontally on stiffer substrates, and vertically on softer substrates. Two crucial markers of colony growth and maturation are the transition of the colony from planar to 3D structure, also referred to as mono-to-multi layered transition (MTMT) \cite{You2019, Berne2018}, and the colony area doubling time, which depict how fast colonies are spreading on surfaces as they grow. By plotting the length of the colonies in terms of their major and minor axes close to the MTMT transition time, a significant variation can be observed when the substrate stiffness is altered. With increasing Young's modulus (Fig. \ref{fig:FIG1_MAIN}(D)), the MTMT occured faster than that for colonies growing on harder substrates. A related trend is depicted by the colony doubling time statistics (Fig. \ref{fig:FIG3_MAIN}(A)), which indicates that over time, colonies growing on harder substrates spread faster. Thus, it is clear that over time, significant structural variations appear in growing bacterial colonies depending on the substrate compliance.  

\subsection*{Substrate stiffness impacts cellular organization in nascent biofilms}
A natural question then is to understand the underlying factors driving the contrasting observations. Is it asynchronous growth and division at the individual scale or is it organizational variation, i.e., the way cells are arranged within the colonies, that determine the observed variations across different substrates ? 

\begin{figure}
\centering
{\includegraphics[width=5.5 in]{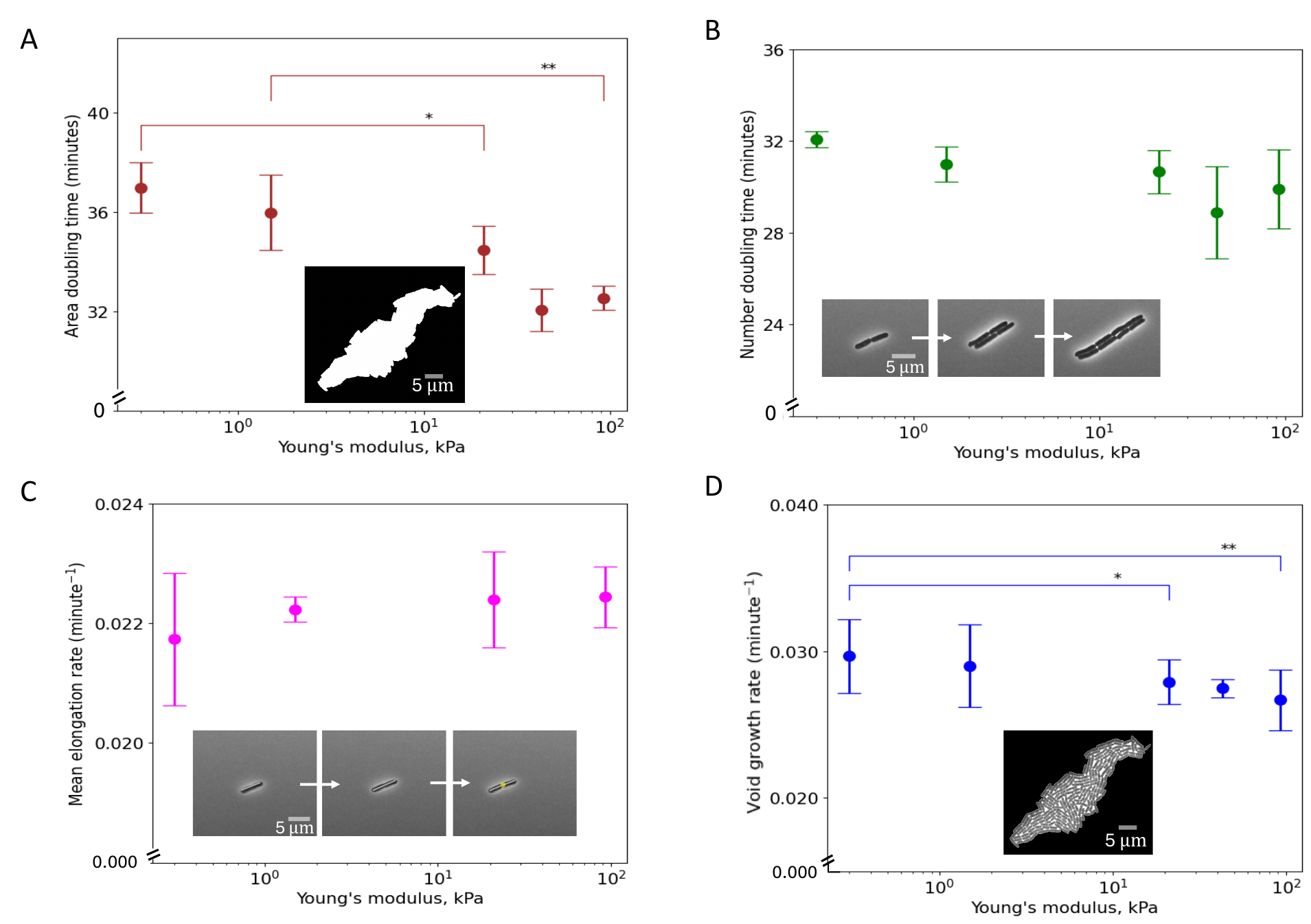}}
\caption{\textbf{Substrate stiffness impacts the growth of bacterial colonies.} Various doubling times are reported here: (A) area doubling time, (B) cell number doubling time, (C) Mean elongation rate of cells, and (D) Growth rate of voids in the colony. Comparisons are made to check the variation in values when stiffness of the substrate is changed. A significant difference in area doubling time and void growth rate values is observed when the Young's modulus varies from from low to high values. Data represent at least three biological replicates per stiffness condition for statistical analysis. Asterisks  corresponds to a specific level of significance: two sample t-test, p-value$<$ 0.05 ($^*$); and p-value $<$ 0.01 ($^{**}$).}
\label{fig:FIG3_MAIN}
\end{figure}

To understand this, we first map the population doubling time. Interestingly, this shows little variation across the different substrates (Fig. \ref{fig:FIG3_MAIN}(B)), as also also indicated by the statistics of the cellular elongation rates (Fig. \ref{fig:FIG3_MAIN}(C)), which likewise show no significant variation across the different cases. Thus, at the level of single cells, we do not observe a statistically significant difference between different cases, indicating that the cellular growth and division are largely unaffected by substrate variations considered here. Notably, some previous studies, for instance on motile bacterial species, have also observed similar behavior \cite{Gomez23}. Thus, colony-scale growth variations which we observe are likely driven by organizational and geometric variations in the expanding bacterial colonies. To probe this, we next focus on the impact of substrate stiffness on the cellular organization in colonies and the overall colony geometry.

\begin{figure}
\centering
\includegraphics[width=\columnwidth] {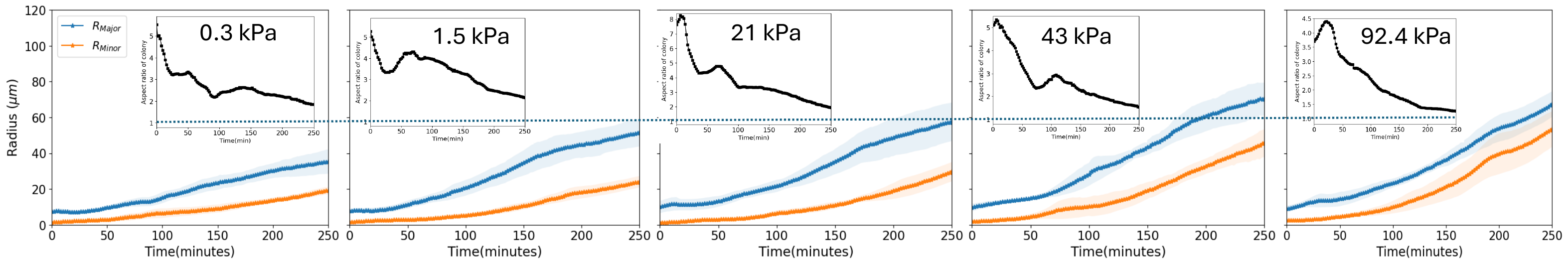}
\caption{\textbf{Colony-scale geometric features.} The major (blue) and minor (orange) axes are plotted as function of time, for the  substrates with increasing Young's modulus as mentioned in inset. Inset also shows the corresponding times series of colony aspect ratio in all cases.}
\label{fig:FIG2_MAIN}
\end{figure}

Cell arrangement in bacterial communities can depend crucially on environmental factors including temperature as well as local settings, for instance, spatial confinements and fluid flow \cite{Nuno2023, Rani2024, Jingyan2021, Hartmann2019}. For us, the question of interest here is how does substrate stiffness, and consequently, cell-substrate interactions affect the growth dynamics as well as the way cells self-organize into microdomains \cite{You2018}.  For this, we start at the colony-scale, where we observe that the overall geometry of the colony shows interesting variations under different cases in consideration here. While the major and minor axes of the colonies show similar statistics over short timescales, they start diverging over longer times for different stiffness conditions (Fig. \ref{fig:FIG2_MAIN}). The growth of the major axes shows an increasing tendency as the concentration of agarose in the substrates is increased. However, this tendency stalls for substrates with higher agarose concentration. Indeed, colonies growing on substrates having $2\%$ and $3\%$ agarose concentration show nearly the similar growth trends. On the other hand, the minor axis of the colonies shows similar growth statistics for the softer substrates, but then as the agarose concentration is increased, it shows more robust growth, and in fact, approaches quite close to the major axes in the case of the substrate with the highest agarose concentration ($\sim 3\%$ agarose concentration). This also suggests that colonies growing on stiffest substrates show a more isotropic circular shape, as revealed by the time lapse imaging data (Fig. \ref{fig:FIG1_MAIN}), and confirmed by the colony-scale aspect ratio, i.e., the ratio of the two axes (Fig. \ref{fig:FIG2_MAIN}).

Next, we zoom into the local cell-level arrangement, corresponding to various substrate stiffness. As cells in growing colonies are constantly being pushed around due to growth-induced activity \cite{You2018, Dell2018}, diverse features can be extracted, and used to study their spatio-temporal arrangement, including microdomain size distribution and local angular and polar distributions \cite{You2018, Shimaya2022}, as well as the formation of genealogical enclaves and their intermixing characteristics \cite{Rani2024}. Here, our aim is to understand the overall size distributions for colonies growing on substrates with different stiffness. We quantify the void formation within the colonies, signifying the empty pockets in the colony which form when cells undergo growth-induced redistribution (Fig. \ref{fig:FIG3_MAIN}(D)). We capture a a clear trend, with a higher void growth rate in case of softer substrates, whereas a lower void growth rate is observed for stiffer substrates. This exhibits that voids are in general larger and grow faster in colonies growing on softer substrates. Overall, this suggests that the packing efficiency of cells within growing colonies can be tuned by substrate compliance, and could be leverage as a metric for predicting relative stiffness of different substrates.

\begin{figure}
\centering
\includegraphics[width=\columnwidth] {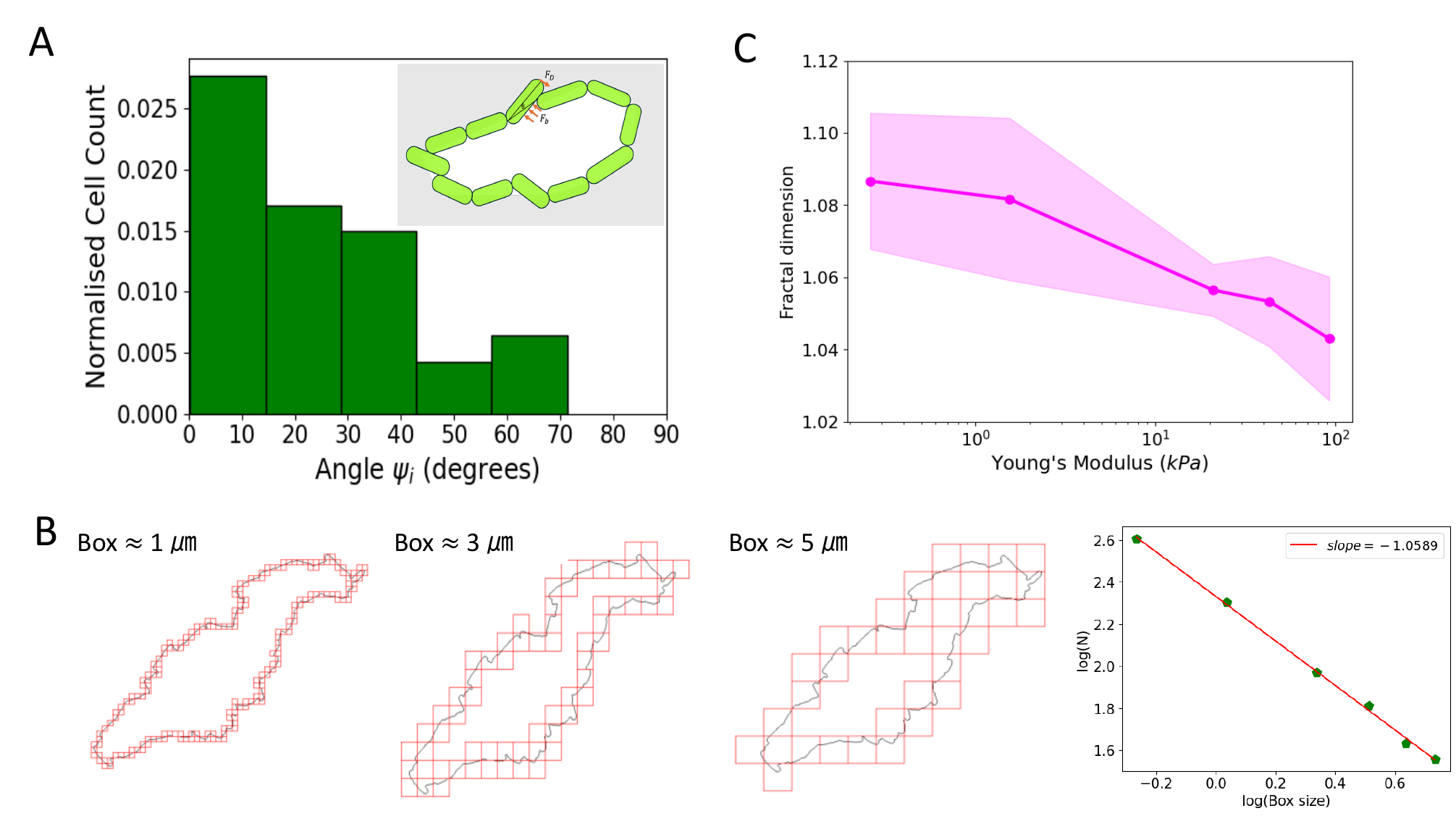}
\caption{\textbf{Boundary roughness of the growing colonies quantified by fractal dimension.} (A) Histogram of the acute angle between the boundary tangent and bacteria orientations. Most cells tend to align with the tangent of the colony boundary. However tangential order breaks at the colony boundary due to imbalance between growth-driven expansion forces (due to the expanding colony) and the opposing drag force from the substrate, leading to orientational disorder at the boundary. High fractal at softer substrate (Inset). (B) Box counting algorithm for computing the fractal dimension of the colony boundary: the colony boundary coordinates are computed and the image is then divided into grids, with grid size progressively increased, and the number of boxes intersecting the colony boundary counted. The fractal dimension in each case is computed by finding the slope of logarithm of the number of boxes containing part of the boundary and logarithm of the box size. (C) The fractal dimension is plotted (mean and standard deviation), for colonies at around transition time for increasing Young's modulus of the LMP agarose substrate.}
\label{fig:FIG4_MAIN}
\end{figure}

\begin{figure}
\centering
\includegraphics[width=5 in] {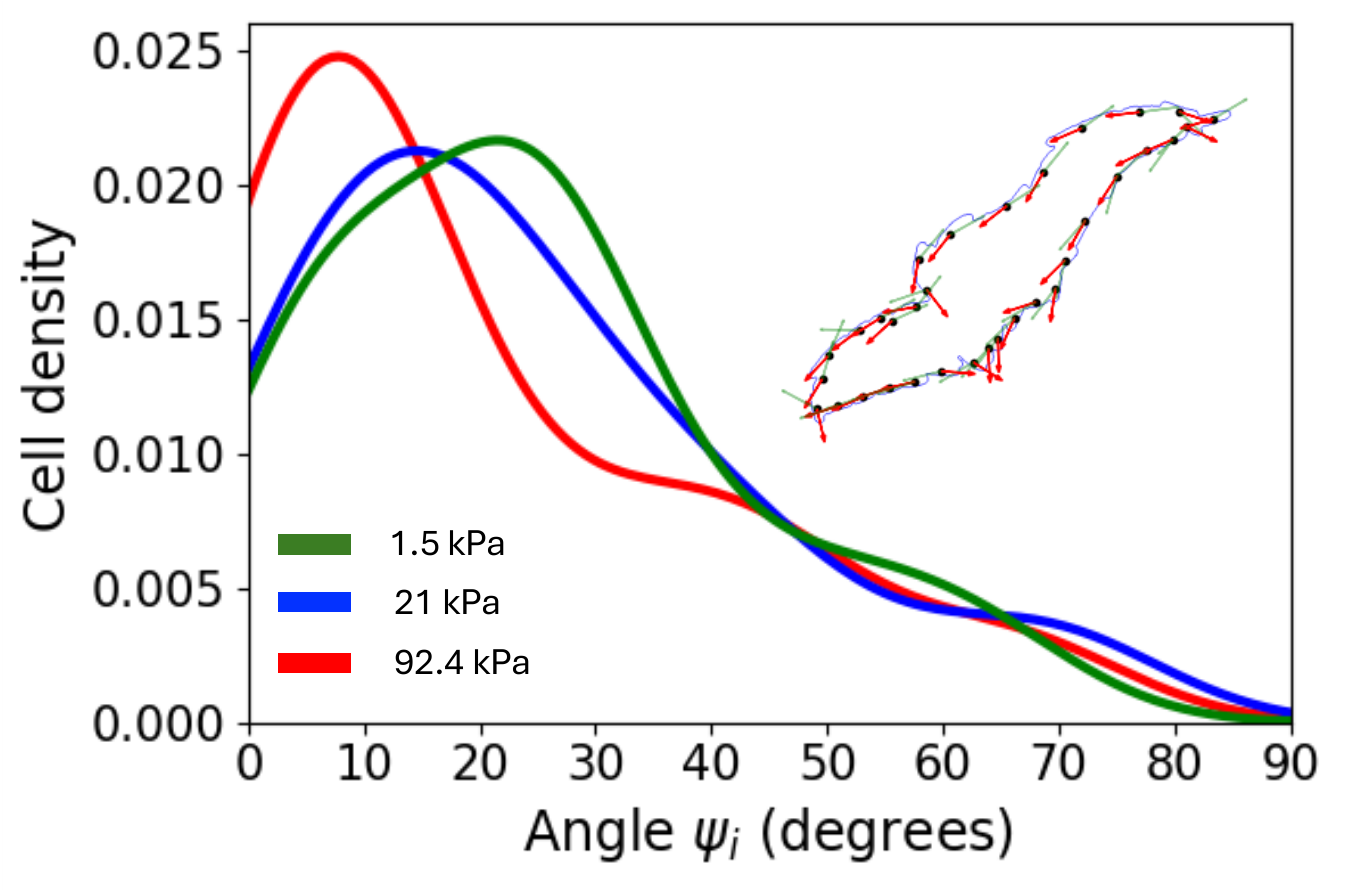}
\caption{\textbf{Cell alignment at the colony boundary.} The probability density of the acute angle between the boundary tangent (green arrows) and the orientation (red arrows) of bacterial cells (black centroids) reveals that most cells tend to align tangentially with the colony boundary across all conditions. However, substrate stiffness influences the extent of this alignment: cells growing on stiffer substrates exhibit a higher degree of tangential alignment compared to those on softer substrates.} 
\label{fig:FIGS31}
\end{figure}

The interactions between cells and substrates are arguably maximised at the boundary of the colony, with the colony actively pressing its way forward to expand. The moving bacterial front may be impeded by substrates, particularly the soft ones, with the boundary functioning as an "active-passive interface" \cite{Zhang2022}. This gives rise to striking interfacial behavior, including a proliferation of topological defects closer to the outer periphery \cite{Sengupta2020, Rani2024, Amin2016}. Consequently, we hypothesise that the substrate compliance will have a considerable impact on the cell arrangement at the colony boundary. We test this hypothesis by quantifying the local orientational order at the colony boundary. We observe that cells on the boundary are largely aligned tangential to the boundary (Fig. \ref{fig:FIG4_MAIN}(A)), as has been noted in interfacial regions in multiple contexts \cite{You2018, Amin2016, Franco2017}. However, disorder occurs when some cells are pushed out of the tangential configuration and adopt orientations angular to the cell boundary, resulting in a characteristic "rough" colony boundary. We perform fractal analysis of the boundary (see Fig. \ref{fig:FIG4_MAIN}(B) and Materials and Methods), allowing us to go beyond the Euclidean geometric methods typically used for 
morphological analysis. Fractal analysis, on the other hand, has been increasingly used in multiple reports to analyze and quantify morphological complexity of cells and their aggregations \cite{Porter91, fractalbook}. In our case, the disorder in the orientation of cells at the colony boundary results in an oscillating boundary curve, whose "magnitude" of oscillation and consequently, the degree of orientational disorder, can be captured by the fractal dimension of the boundary curve. Our data shows that fractal dimension of the boundary curves of colonies decreases with increasing substrate stiffness (Fig. \ref{fig:FIG4_MAIN}(C)). The results indicate that the higher fractal dimension, capturing a higher degree of disorder at the boundary, is observed for colonies on soft substrate. Fractal dimension closer to 1, as seen in the case of hard substrates, suggest a smoother curve with most of the cells aligned tangentially with the colony boundary. This is also indicated by the distribution of the angular difference (between the colony boundary tangent and cell orientation at MTMT, Fig. \ref{fig:FIGS31}), which peaks at very low angles for colonies growing on stiffer substrate. Thus, we can conclude that a larger number of cells and consequently a higher order of orientational disorder occurs in colonies growing on softer substrates.

\subsection*{Structural complexity of nascent biofilms depends on substrate stiffness}

As noted, bacterial colonies initially grow in a planar fashion and then transition to layered 3D structures, which over time spread and grow vertically akin to a spherical cap, a phenomena that is widely conserved across bacterial species and morphotypes \cite{Jingyan2016, Hwa2019, Yunker2024}. This transition, referred to as mono-to-multi layer transition (MTMT), occurs when the active in plane stresses in the colony interior become high enough to overcome restoring forces and force cells to climb on top of the base layer \cite{Wingreen2018,You2019}. We have reported above that substrate stiffness affects this key feature of colony maturation, and here we carry out a mechanistic analysis of the same.

\begin{figure}
\centering
\includegraphics[width=5 in] {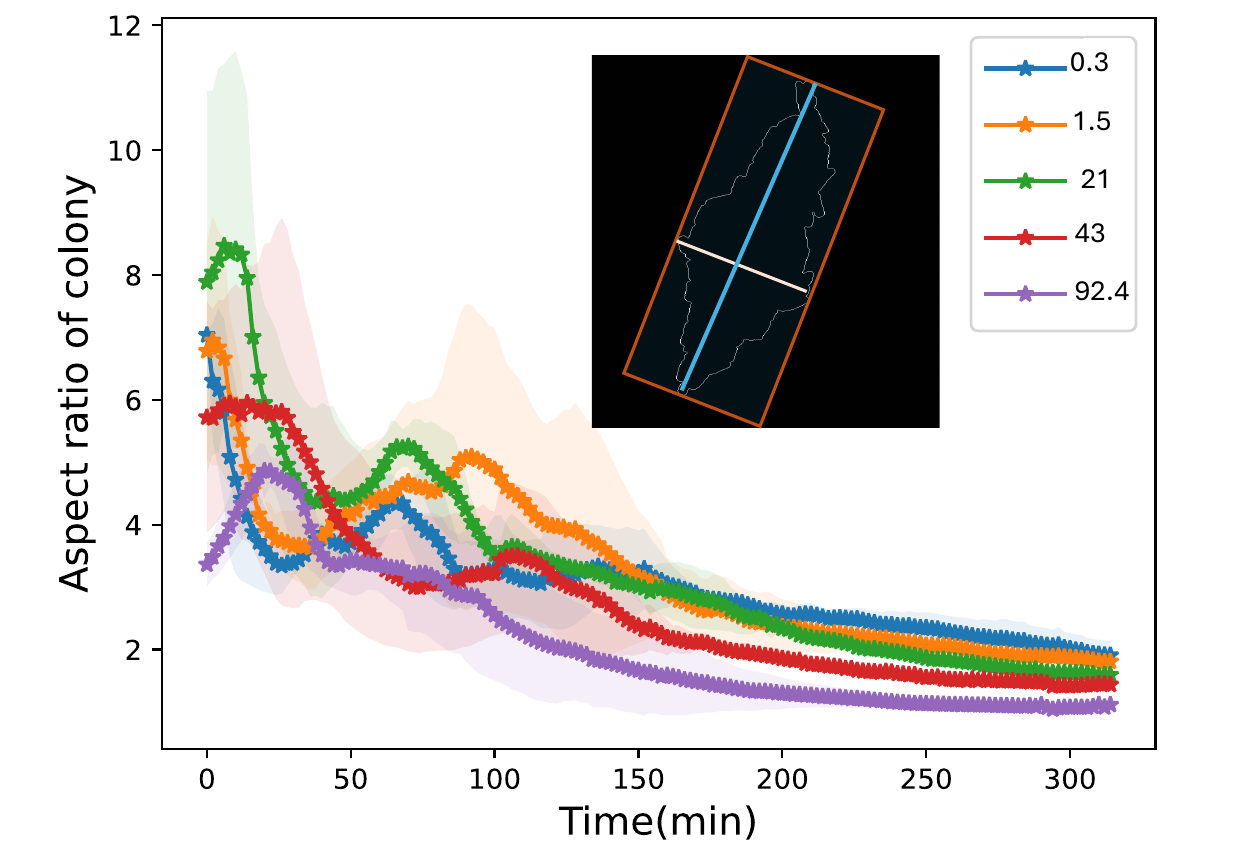}
\caption{\textbf{Shape anisotropy of the colonies.} Colony aspect ratio is plotted as a function of time of colony growth for various substrates as labeled by their Young's modulus. Inset shows the bounding box on the colony contour to compute the major and minor distance of the colony. The mean (solid line) and standard deviation (shaded region) are calculated from at least three biological replicates for each stiffness condition.}
\label{fig:FIGS1}
\end{figure}

Specifically, how do the planar to 3D transition attributes, and following the transition to the 3D geometry, the new morphology depend on the substrate stiffness? Visual inspection of the time-lapse images of growing colonies (Fig. \ref{fig:FIG1_MAIN}(A)) suggests striking differences in the 3D structural geometry of colonies growing on soft versus hard substrates. For the softest substrates considered here, the second layer adopts a similarly sized and shaped structure as the initial monolayer, making them optically indistinguishable. However, for colonies growing on harder substrates, a marked difference in the layers can be observed: the layers are relatively circular in shape, and the 3D geometry vis-a-vis the initial planar geometry, is clearly distinguisable, owing to the visual contrast across different layers. For colonies growing on low agarose substrates, i.e., soft substrates, the second layer is spread out almost completely over the base layer, while for high agarose case, layers growing after colony transition into 3D structure have a much smaller spread compared to the base layer. In general, for all cases, colonies tend to adopt morphologies which bacome circular with time, particularly after the transitioning to multi-layered structures (Fig. \ref{fig:FIGS1}). This shows that substrate properties modulate 3D geometry of colonies from the inception of layer addition after the planar monolayer growth, while it was explored in past studies that large scale 3D structures of bacterial colonies show dependence on substrate properties \cite{MATSUSHITA1990, Geisel2022, Jingyan2023}. Notably, for large 3D colonies, similar observations have been made, where as substrate stiffness increases, it has been seen that colonies tend to adopt more packed and flattened geometries while for softer substrates, colonies adopt loose, dome-like geometries \cite{Jingyan2023}, similar to our observations here, and as suggested by our void analysis (Fig.\ref{fig:FIG3_MAIN}(D)).

\begin{figure}
\centering
\includegraphics[width=\columnwidth] {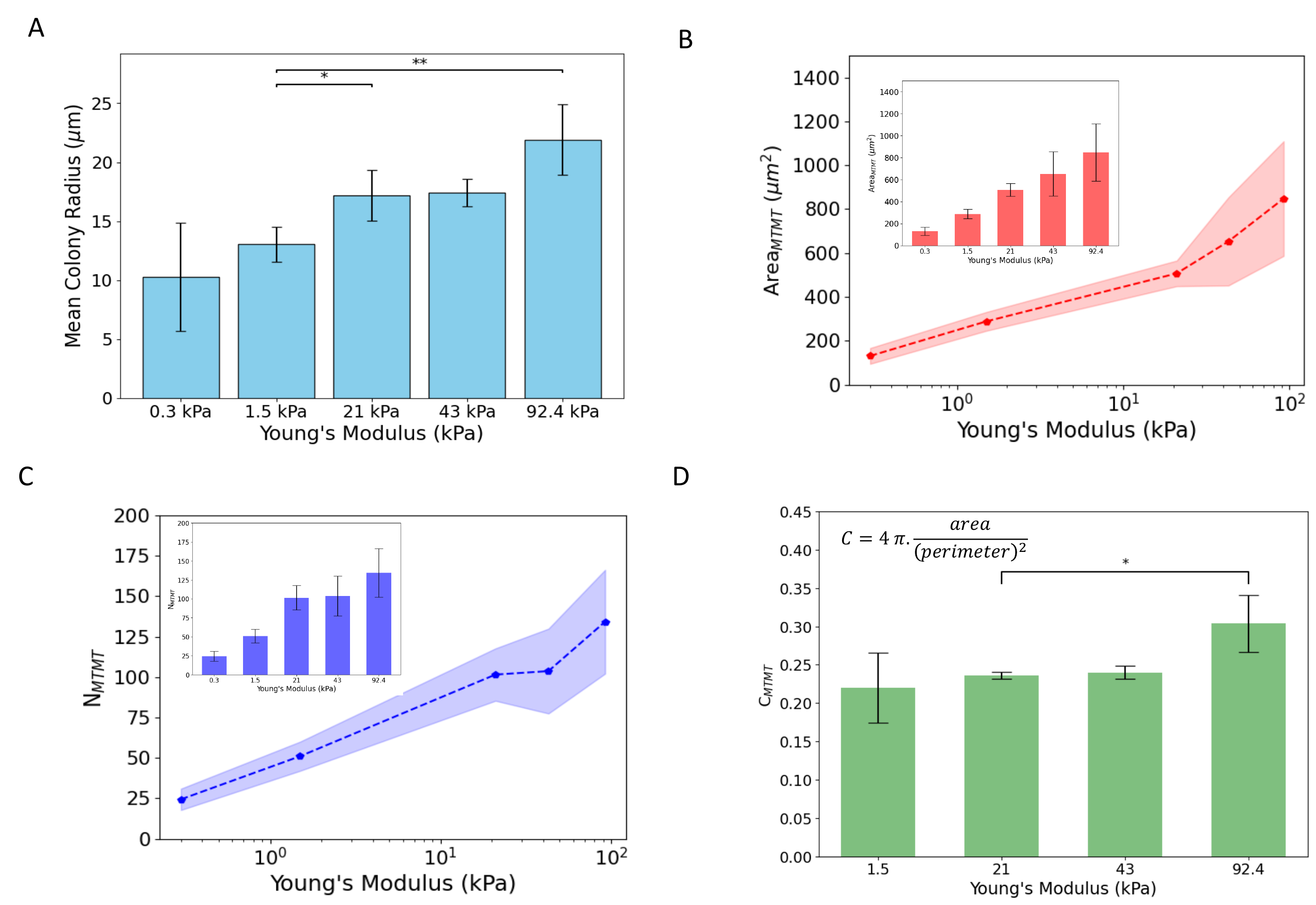}
\caption{\textbf{Growth dynamics and geometric parameters as the colony transitions to the multilayered morphology.} Variation of the (A) mean colony radius, (B) colony area, (C) cell number count, and (D) circularity (measures the compactness of the colony morphology) at transition time from single to multi-layered morphology, as substrate stiffness varies. A significant difference in mean colony radius, colony area and number of cells at the MTMT is seen when Young’s modulus varies from 1.5 kPa to high values 92.4 kPa (two sample t-test, p-value$<$ 0.01). Circularity values also display variation with the Young's modulus, between 21 kPa to 92.4 kPa (p-values $<$0.05).}.
\label{fig:FIG5_MAIN}
\end{figure}

\begin{figure}
\centering
\includegraphics[width=6 in] {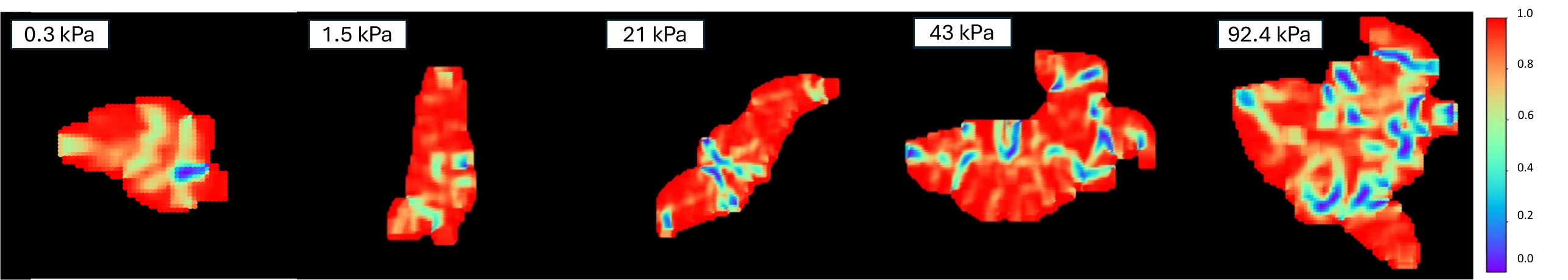}
\caption{\textbf{Nematic order parameter for colonies growing on substrates with different Young's modulii.} Color map of distribution of values of nematic order parameter S in shown for a representative case for colony at MTMT time growing on substrates with increasing Young's modulus from left to right (MTMT time of bacteria colonies: $t_c \approx 130$ min for 0.3 kPa;  $t_c \approx 160$ min for 1.5 kPa; $t_c \approx$ 200 min for 21 kPa and $t_c \approx 235$ min for 92.4 kPa substrate.) } 
\label{fig:FIGSOP}
\end{figure}

Colonies growing on softer substrates exhibit an early onset of the mono-to-multilayer transition, as compared to those growing on harder substrates (Fig.~\ref{fig:FIG1_MAIN}). To understand this, we probe multiple spatio-temporal characteristics related to 3D transition in the various cases. First, we compare the size of the colony at around the MTMT onset time. We observe that colonies growing on softer substrates attain MTMT at smaller colony sizes (measured in terms of colony radius, colony area as well as in terms of the number of cells), as compared to substrates with higher agarose concentrations (Fig. \ref{fig:FIG5_MAIN}(A, B, C)), as also reflected in the distribution of the orientational order parameter S, which shows a wider distribution of values at MTMT (Fig. \ref{fig:FIGSOP}, see also Ref. \cite{Rani2024}). Numerically, the colony size at the transition increases with increasing substrate stiffness and the difference is similar when comparing area and number. For instance, colonies on the softest substrates considered here are around 4-5 times smaller at MTMT as compared to the colonies growing on the hardest substrate considered here, both when comparing area and number of cells. It is pertinent to note here that, as observed above, the colonies in different cases show similar areal growth statistics, initially (Fig. \ref{fig:FIG1_MAIN}(B,C)), which in fact is related to their structural transition to a 3D structure. Indeed, colonies growing on soft substrates spread at a similar rate as those growing on harder substrates till around the time they transition to a multi-layered structure, after which their spread relatively slows down. Further, as noted, the colonies on all the substrates get more circular with time, but circularity around transition time also shows a dependence on substrate properties, showing an increasing trend with substrate stiffness (Fig. \ref{fig:FIG5_MAIN}(D)).

It might be worth noting that an interesting dichotomy is observed here: since the transition to 3D morphology on softer substrates occurs faster than those growing on harder substrates, this seemingly suggests a higher magnitude of internal stresses building up in these colonies, promoting early onset of MTMT. This is likely due to a more pronounced impediment to the colony spread on the substrate (resulting in effective drag). However, we recall that boundary orientational disorder is also higher for colonies growing on softer substrates, which suggests that cells get perturbed more easily out of their position on softer substrates. To clarify this, we need further understanding of the biophysical underpinnings of colony growth on substrate, integrating information at both the colony as well as the cellular scales.

\begin{figure}
\centering
\includegraphics[width=\columnwidth] {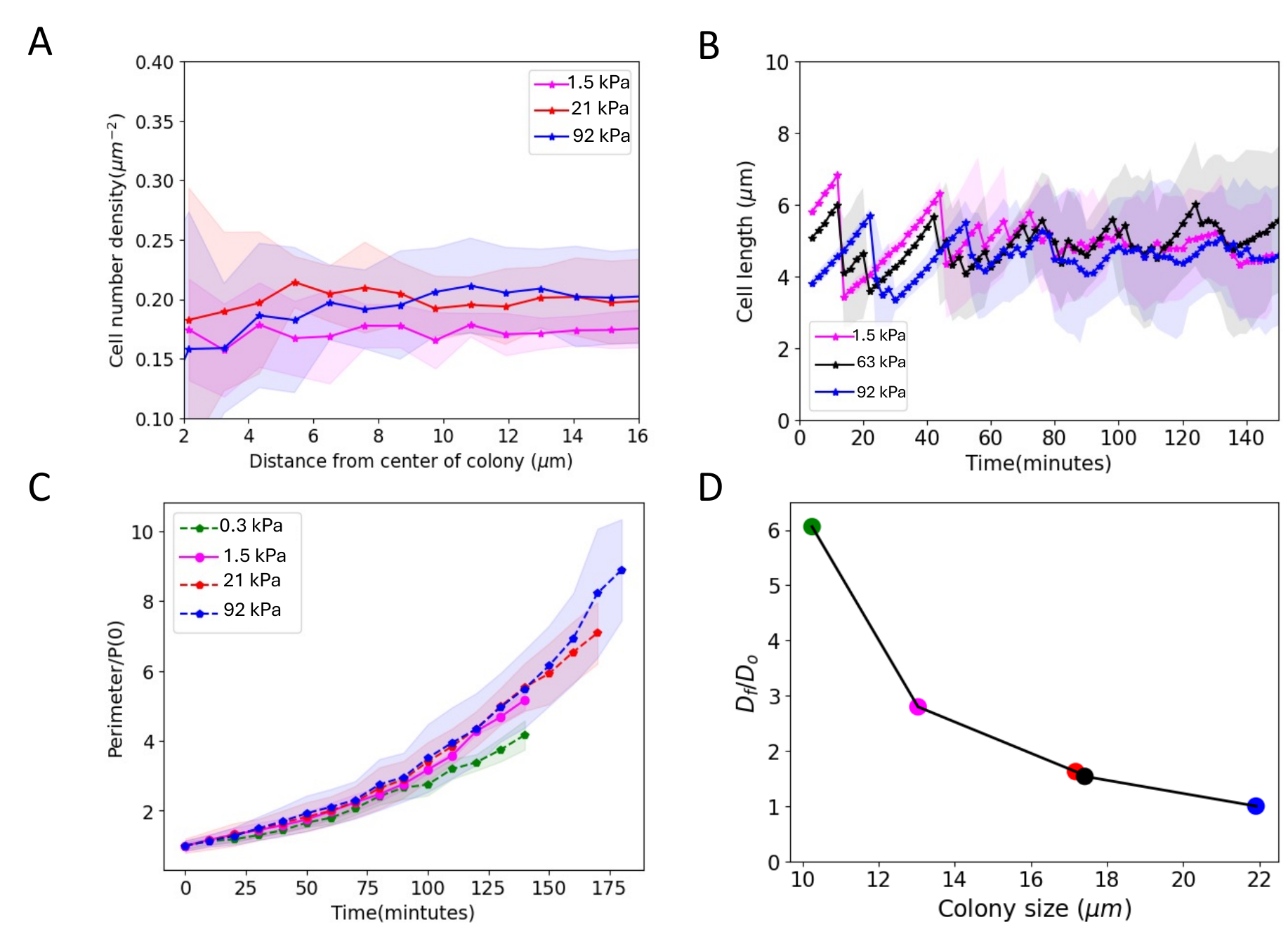}
\caption{\textbf{Spatiotemporal changes in the colony due to the variation of the substrate stiffness.} (A) Spatial variation in the cell number density is plotted as a function of distance from the colony center, (B) Mean cell length of cells in the colonies is plotted as a function of time, and (C) Normalized colony perimeter is plotted with time for different substrates with varying Young's modulus. (D) Estimated drag parameter $D_f$ (normalized by the drag parameter for the case of colonies growing on  hardest substrate, having 3$\%$ LMP agarose concentration) is plotted as a function of colony size at the transition to multilayered growth. Here, each point correspond to different Young's modulus values.}
\label{fig:FIG6_MAIN}
\end{figure}


\subsection*{Substrate-mediated drag forces impede colony expansion on soft substrates}

We develop a simple biophysical model of cells growing on a soft substrate to shed light on the possible mechanisms through which cell-substrate interactions modulate cellular organization and large-scale structures within the microcolonies. Our model allow to qualitatively understand the biophysical underpinnings of the observed phenomena, including the dichotomy in degree of orientational disorder at colony edge as well as the early nucleation of multi-layered colony structure, as detailed above. 

We hypothesise two primary modes of cell-surface interactions in a developing bacterial colony: on the one hand, the drag forces which impede the displacement of cells on the substrate, and on the other hand, adhesive forces, which are specific to the interactions between the cell surface and the substrate, which may or may not be actively modulated by the cells, depending on external conditions. In our case, the timescale of interest is the early stages of colony maturation, centered around the attainment of layered configuration. At this timescale, in our previous work, it had been reported that very little variation can be observed in adhesive forces between cells and substrate in the various cases \cite{Rani2024}. So, here we focus on drag mediated colony-substrate interactions as they spread on it.

\begin{figure}
\centering
\includegraphics[width=4in] {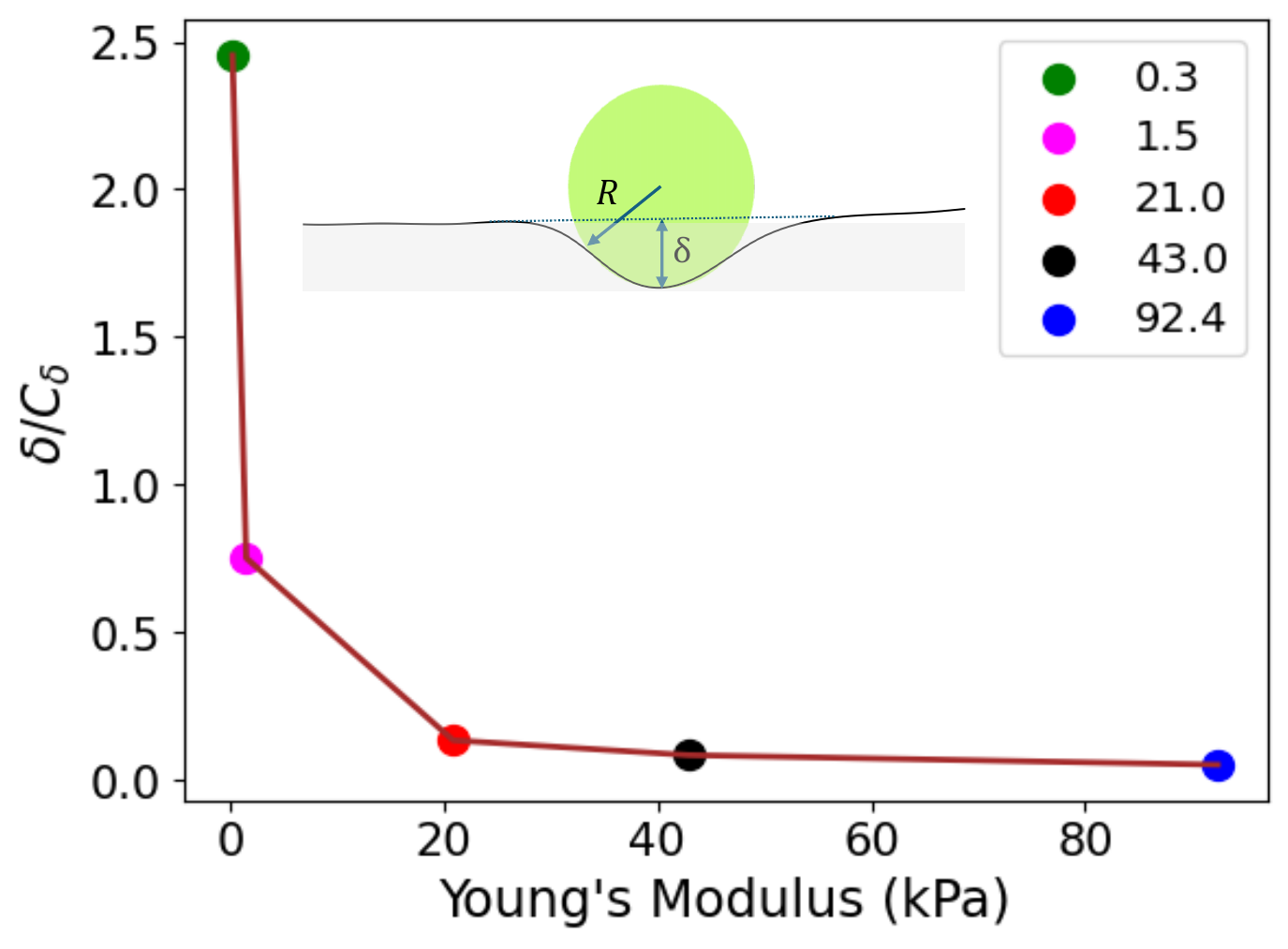}
\caption{\textbf{Emergence of effective drag forces on growing colonies.} Relative sinking depth estimated for colonies growing on the various substrates, with dots indicating the Young's modulus values. (Inset) Illustrative sinking of colony into substrate, with depth estimated using Hertzian contact.}
\label{fig:FIG7}
\end{figure}

As a first step, we estimate how the difference in LMP agarose concentration results in the relative difference in drag forces experienced by the growing colonies in the respective substrates, utilizing our colony size data at transition time (Fig. \ref{fig:FIG5_MAIN}). It has been shown that planar to 3D transition occurs when internal stress on a cell exceeds a critical stress $\sigma_c$, which depends on the size distribution of cells in the colony and the magnitude of the adhesive force acting perpendicular to the planar colony to oppose extrusion \cite{You2019}, which in our case are the same for all cases, as shown in our previous work for cell-substrate adhesion and in (Fig~\ref{fig:FIG6_MAIN}A) for cell length distribution in the colonies. For simplicity, we will take the colony size at which internal stress at the center exceeds $\sigma_c$ as a proxy for average colony size at transition.

We first consider a simple 1D chain model of bacteria, length $L$ that spreads across $-L/2$ to $L/2$, to quantify how the relative differences in colony size around the transition time, and indicate the relative difference in the drag force that is experienced by the colonies as they spread on the various substrates. The stress evolution in the 1D colony follows: 

\begin{equation}
       \partial_x \sigma_x + D_fv_x=0
\end{equation}

\noindent where $\sigma_x$ denotes the growth-induced normal stress along the $x$ axis. The colony density evolves as

\begin{equation}
      \partial_t\rho_c+\partial_x(\rho v_x) =k_c\rho 
\end{equation}

\noindent where $\rho_c$ is the density, $v_x$ denotes the direction of velocity along the $x$, and $k_c$ is the growth rate. $D_f$ is a drag parameter, that quantifies the scale at which colony expansion is impeded due to the substrate interaction. With $\rho_c$ remaining constant throughout the colony spatio-temporally (Fig. \ref{fig:FIG6_MAIN}(B)), and taking free boundary conditions $\sigma(-L/2)=\sigma(L/2)=0$, we get

$$\sigma_x (x)= \frac{k_cD_fL^2}{8}\bigg(1-\bigg(\frac{2x}{L}\bigg)^2\bigg)$$

where $L$ denotes the size of the colony. Stress is maximum at $x=0$ and so, for the critical stress value $\sigma_c$, we can derive colony size at transition as

\begin{equation}\label{eq:Dragf}
    L_c=\sqrt{\frac{8\sigma_C}{k_cD_f}}
\end{equation}

As noted above, colonies spread at a similar rate for all cases initially, so we can derive a relative comparison between values for the drag parameter $D_f$ for different cases, by comparing the colony sizes around transition time and using Eq. \ref{eq:Dragf} (Fig. \ref{fig:FIG6_MAIN}(D)). We see that drag is much higher in case of colonies growing on softer substrates, indicative of a higher propensity to impede colony expansion. This results in quicker transition of bacterial colonies to multi-layered structures. 

To further understand the mechanical reason for this, we now analyse the drag force in case of growth of a 2D colony. Recall the Rayleigh's drag equation

\begin{equation}\label{eq:drag force}
    F_D=\frac{1}{2} \rho C_d v^2 A
\end{equation}

where $F_D$ is the drag force experienced by the moving object, $\rho$ is substrate density, $C_d$ is the drag coefficient which may depend on factors including velocity of the moving object, shape of the moving object and substrate properties, $v$ and $A$ denote the velocity and the cross-sectional contact area of the moving object. As the drag force is much higher for softer substrates, Equation \ref{eq:drag force} suggests that a key role in this regard is played by the cross-sectional area $A$. The cross-section area here is the area of the object that faces resistance while moving, so in the case of the colony growth, the area is related to the expanding colony front and is determined by the colony perimeter and the depth to which it sinks into the substrate. Note that the necessity of considering 2D colony expansion is vis-a-vis the simple 1D chain model is clear now, as a 1D colony expands only in two directions (front and rear of the chain), so the area will remain constant, equal to twice the area of cell that is sunk into the substrate in its front end. However, for the 2D colony expansion, the colony front is constantly increasing, which leads to a rapid increase in the contact area with time. To estimate the evolution of the contact depth, we consider the 2D Hertzian  contact problem where we assume a spherical shape for simplicity, in which case the indentation depth for a sphere of radius $R$ is (Fig. \ref{fig:FIG7}): 

\begin{equation}
    \delta \approx (\frac{9F^2}{16E_{cs}^2R})^{\frac{1}{3}}
\end{equation}

\noindent where $F$ is the normal force of the cell mass and $E_{cs}$ is the combined cell-substrate elastic modulus \cite{Landau86}. The effect of substrate stiffness is encoded in the combined cell-substrate elastic modulus $E_{cs}$ which is given as 

\begin{equation}
    E_{cs}=\frac{4E_cE_s}{3(E_c+E_s)}
\end{equation}
where $E_c$ and $E_s$ denote the elastic modulus of cell and substrate respectively and the Poisson ratio have been taken to be equal to 0.5 for cell and substrate \cite{Rene2024}. However, as $E_c$ values, given by the elastic modulus of bacterial cells is around $\sim 20-50 MPa$ \cite{Yao1, Tuson2012}, so $E_c>> E_s$, giving $E_{cs}=4E_s/3$. Thus, estimating the relative sinking depth for similar sized colonies on the various substrates (Fig.\ref{fig:FIG7}), we observe that the contact depth varies inversely with substrate stiffness. Now, the contact area for a single cell of length $l$ is then given as $\sim l\delta$ (for a typical cell, this will be $\sim 0.002-0.03\mu$m$^2$), while for a colony with an expanding front of perimeter $P$, the contact area will be $\sim P\delta$ (for a colony of 100 cells of radius 15$\mu m$, this will be $\sim 0.5-3\mu$m$^2$). As noted, this encapsulates the difference in the 1D vis-a-vis the 2D case, with the exponential growth in the 2D case creating a key difference in the way drag forces act on growing bacterial colonies (Fig. \ref{fig:FIG6_MAIN}(D)). Note, we have not considered the relative role of differences in microstructure of LMP agarose substrates as a function of concentration, which can also play a contributing role in colony spread dynamics \cite{agaroseProps}.

However, why do colonies growing on softer substrates display a higher "roughness" on colony boundaries and orientational disorder, even though they are subjected to higher drag forces from the substrate which may act to enforce restoring of tangential order at the boundary? To understand this, we analyse the breaking of tangential order by cells at the boundary. In general, cells in the colony boundary are aligned tangentially, occasional cells break this symmetry and adopt angled orientations at the colony-substrate interface. This phenomena can be well-understood as a balance between colony internal stresses and restoring forces at the colony-substrate interface- such instabilities at the boundary are local in nature, caused when the colony is expanding in a given direction, breaking the local symmetry of the tangentially aligned cells at the boundary. We carry out a force analysis for the triggering of such an instability at the colony-substrate border. Say, a tangentially aligned cell is rotated out of position by this force, taking an angle $\theta$ to the colony boundary Fig.~\ref{fig:FIG4_MAIN}(C). The local force from the expanding bulk, which we denote $f_b$, acts to move cells out of tangential orientation while the opposing force due to drag, is acting against the motion of the cell as it rotates to change its orientation. The torque $\tau_b=lf_b\cos\theta$ is opposed by the drag torque, which is given by $\tau_d=lF_D$. Thus, cells will rotate out of the tangential alignment whenever the local bulk forces exceed $F_D$, which is directly proportional to $\rho A$, where $\rho$ is the substrate density and $A$ is the cross-sectional contact area (see Eq. \ref{eq:drag force}). However, the cross-section area is now at the single cell level and is bounded by $l\delta$. So, we can conclude that orientational disorder at the boundary is decided at single cell length scale and in this case the cross-section area is much less. This allows it to be dominated by the bulk forces relatively easily for softer substrates which have lower density, allowing cells to move out of tangential position more easily in these cases. This results in more disordered orientation of cells at the colony boundary. On the other hand, as the colony grows in 2D, the cross-section area under consideration in across the colony perimeter which grows exponentially, so combined with the greater depth of cells sinking into the softer substrates, this results in higher drag force and thus, faster transition into multi-layered colony morphologies. 

\section*{Summary and discussion}

Growth of bacterial colonies on surfaces is one of the most important factors underpinning the resilience and ubiquity of bacteria in a wide variety of ecological settings. Still, a clear understanding of the way bacterial cells interact with soft surfaces while growing into a colony, has not been achieved. By varying the stiffness of substrates, achieved by tuning concentrations of low melting agarose, we focused on understanding bacteria-substrate interactions, specifically to uncover if bacterial colonies during their early stages of development, display any dependence of compliance of the underlying substrates. Our results indicate multiple organizational features, from individuals to colony-scales, were impacted by the substrate interactions. While the colony spread in all cases was similar over the short time-scales, there was sharp divergence as the colonies got bigger, with those growing on harder substrates showing a faster and wider horizontal spread. To understand this further, we analyzed the morphology of these colonies in detail, revealing that the growth and division dynamics of cells growing on substrates with different compliance remain statistically comparable, however, the self-organization of cells within these microcolonies is impacted. We observed a greater propensity of void growth in colonies on softer substrates as compared to those growing on harder substrates. In fact, colonies showed a relatively higher degree of orientational disorder at the colony boundaries for softer substrates, as revealed by the fractal analysis of the colony boundaries. Further, analyzing the transition of the colony from planar to multi-layered 3D morphologies, revealed that colonies growing on softer substrates transition more rapidly into multi-layered structures, i.e., expand vertically rather than horizontally (for the ones growing on harder substrates). Upon such transition, the 3D geometry of the colonies also showed marked differences, with colonies growing on harder substrates showing a clear demarcation between the layers and a circular, isotropic layout, while colonies growing on softer substrates display texturally heterogeneous structures, with the second layer closely following the spread of the base layer. Biophysical modeling then revealed the key role of effective drag forces experienced by the colonies as they spread on the different substrates. Drag forces acting at the individual-scale on the one hand and the colony-scale on the other hand, promotes faster transition to multi-layered structures for colonies growing on softer substrates while also result in higher orientational disorder at the colony boundary. Similar insights into stochastic and deterministic spreading dynamics have previously been reported using agent-based and continuum models \cite{Rana2017}.

Our work clearly delineates the wide-ranging impact that substrate properties may have on the geometry and self-organization of bacterial species across multiple scales. This may be of particular significance for understanding the structure and distribution of the microbiomes associated with humans, which typically span a diverse range of surfaces and geometries. The study of cellular organization in colonies which spread across interfaces, for instance across tissues and implants in the human body, can play a predictive role in understanding implant failure. Further, here we have concentrated on the effect of substrate compliance on the development of bacterial colonies, with our biophysical model showing how colony growth can be impacted by mechanical interactions and deformations caused in the substrate by colony growth. A detailed mechanistic model of the phenomena reported here is currently bein developed, and will be reported elsewhere. Initial studies have also demonstrated that bacteria are capable of causing significant deformations in host surfaces that can induce a mechanical model of infection \cite{Persat2020}. Indeed, bacterial colonies can change their material properties under mechanical stresses, but this change also depends on substrate properties like agar concentration \cite{Kochanowski2024}. Hence, the mechanical cross-talk between hosts and microbial agglomerations is shaping up to be a promising and potentially illuminating topic of research \cite{Sengupta2020}.

Finally, in our study here, we have focused on the case of early stages of colony formation, around the time of initiation of multi-layered structures for single founder colonies of sessile bacteria, growing on LMP agarose substrates. Thus, there are important distinctions in our work and some of the past work, which typically have focused on large-scale cell agglomerations comprising many thousands of cells, in several cases, on communities of motile bacteria and on substrates made of different materials. For instance, in many cases, growth has been observed to slow down as the substrate stiffness increases \cite{Saha2013, Song2014, Little2019, Yan2017} though in some cases it has been seen indeed to increase with stiffness \cite{Asp2022}. In fact, studies in fungal systems have shown that stiffer substrates can promote faster surface expansion, suggesting that mechanical constraints can directly modulate growth rates \cite{Yang2024}. For large colonies, nutrient availability can play an important role in determining colony behavior \cite{Allen_2019} but at the scale of colonies that we have studied here, nutrient availability is clearly uniform across the colony. Further, for motile bacterial species, even important biophysical considerations might be different- for instance, in our past work \cite{Rene2024}, we had shown how the adhesive behavior of \textit{C. okenii} can be very different from nonmotile \textit{E.coli} bacteria. Thus, an important avenue for future work is to understand the colony behavior vis-a-vis colony size, to map out transitional behavior related to colony spread and growth, to understand how cell motility impacts such phenomena and the biophysical underpinnings of motility induced differences in colony dynamics. 

\section*{Methods}

\subsection*{Cell culture and experimental protocol}

We used non-motile strain of Gram negative bacteria {\it{E.coli}}, namely NCM3722 delta-motA. Single colonies were initially streaked on LB-agar plates and incubated at $30 ^\circ$ C. Individual colonies were picked using a sterile inoculation loop and transferred to liquid LB medium in a shaking incubator set at 170 rpm. Cultures were grown to late exponential phase, as monitored by optical density (OD), and then diluted 1:1000 into fresh LB medium. These cultures were grown for an additional 1.5 to 2 hours to reach a mid-exponential growth phase.  

To prepare the LB-agarose gel substrate, low-melting-point agarose (LMP, Preparative Grade for Large Fragments ($>$ 1,000 bp), Promega, USA) was mixed with LB medium to create an agarose-LB solution. The solution was poured into a Gene-Frame (spacer) adhered to a clean microscope slide. After pouring, the gel was allowed to cool and solidify. Approximately 1 to 1.5 $\mu$ L of bacterial culture was inoculated on the substrate and single cell-to-colony dynamics was observed using time-lapse microscopy. 

The Young's modulus of the LMP agarose gel substrates were measured using atomic force microscopy (AFM), using contact mode and fitted with standard elasticity models, as detailed in our previous work \cite{Rene2024}.

\subsection*{Image acquisition}
The time-lapse imaging protocol, with $2$ minute interval, was designed to capture bacterial growth and colony formation dynamics. LMP agarose concentrations was varied from $0.75\%$ to $3\%$, in the prepared substrates. Multiple regions on the substrate were imaged for every experiment. At least three biological replicates were performed for each substrate stiffness, as well as multiple technical replicates were captured for each biological replicate, covering different positions on the sample. The growing colonies were observed using phase contrast microscopy (Olympus IX83, $60\times$/$100 \times$ oil objective, and imaged using Hamamatsu ORCA-Flash camera), within a temperature-controlled incubator. The experiments were performed at $30 ^\circ$ C to match the growth conditions of the bacterial strain. Single bacterial cells acted as monoclonal nucleation sites, expanding horizontally on the nutrient-rich LB-agarose layer. Initially, the colony expanded horizontally, as a monolayer in two dimensions. Over longer timescales, the bacterial monolayer transformed into multiple layers, thereby developing into a three dimensional structure over multiple generations. Further details on the experimental protocol can be found in our previous studies \cite{Dhar2022, Rani2024}.

\subsection*{Image analysis}
Phase-contrast images were processed using Fiji:ImageJ \cite{Schindelin2012}, Ilastik \cite{Berg2019}, and custom python scripts developed in-house. The image data were pre-processed, involving brightness adjustment and background noise subtraction (via ImageJ), followed by additional noise reduction and filtering using python-OpenCV. A subset of images was used to train Ilastik’s pixel classifier to separate cells from the background, and the classifier was applied to the remaining frames via batch processing, iterating for improved segmentation. Segmentation was performed until the colony transitioned to the third dimension, tracked carefully by monitoring changes in the image contrast within the bulk of the colony. Cell length was calculated as the distance between the centroid and the poles (extreme ends), validated using OpenCV ellipse fitting for the major axis. Colony boundaries were detected using OpenCV’s \textit{findContours} and \textit{flood-filling} routines, allowing us to obtain the boundary coordinates, colony area as well as the perimeter. Void growth rate was determined by establishing the time series of the area of voids (regions within the colony boundary which were not occupied by cells) in the colony and measuring their growth rates. Statistics reported in our analyses and figures were calculated across all replicates.

\subsection*{Computations}

\subsubsection*{Colony-scale geometric features}
\begin{figure}
\centering
\includegraphics[width=\columnwidth] {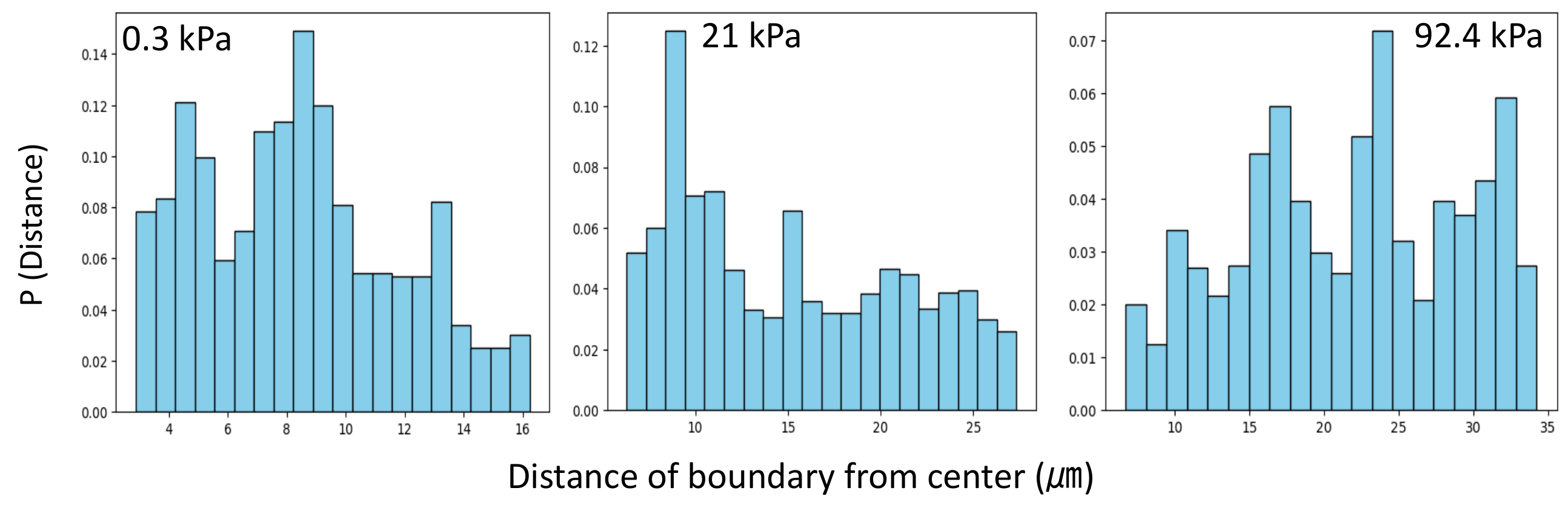}
\caption{\textbf{Boundary distance from the colony center.} Probability distribution of distance from the colony center across a $360^{\circ}$ angle for substrate of 0.3 kPa, 21 kPa and 92 kPa Young's modulus values, at MTMT. The mean value of distance from center ($\sim R_{mean}$) for different cases were $10 ~\mu m$, $17.5 ~\mu m$ and $22.5 ~\mu m$ respectively.}
\label{fig:FIGS3}
\end{figure}

Colony geometry and shape parameters were calculated by fitting a rotating bounding rectangle to the colony boundary and computing its major and minor dimensions, as described in Figs. \ref{fig:FIG2_MAIN} and \ref{fig:FIGS1}. Colony shape was assessed by calculating aspect ratio ($R_{major}/R_{minor}$) and circularity (a measure of compactness, defined as the ratio of the area of the colony to the area of a circle with the same perimeter). Furthermore, for each set of experiments across substrates with different Young's modulii, we extract the distance of the colony boundary from the corresponding center. We present this as the probability distribution of distance from the center of the colony over $360^{\circ}$ angle, as shown in Fig. \ref{fig:FIGS3}.

\subsubsection*{Fractal analysis}
The fractal dimension of the colony boundary was calculated using the box-counting method. First, the grayscale image was converted into a binary image, and the colony boundary was extracted. The box-counting algorithm was applied by dividing the image into grids of varying sizes, and counting the number of boxes which intersected with the boundary. This process was repeated with progressively smaller grid sizes (down to approximately $0.5 ~\mu m$). A plot was generated of the logarithm of grid size (box size) versus the logarithm of the number of boxes which intersect the boundary. The slope of the plot corresponds to the fractal dimension of the boundary (Fig.~\ref{fig:FIG5_MAIN}B).

\subsubsection*{Cell orientation relative to the colony boundary}
Segmented binary images of bacterial colonies were analyzed to study local cell orientation near the colony boundary. Cell centroids within a defined distance from the edge of the colony were extracted. A closed B-spline curve was fitted through these centroids to represent the colony boundary locally. For each cell, the acute angle between its major axis and the tangent to the B-spline at the nearest point was computed. These angles measured the alignment between cell orientation and the local boundary direction. The probability density (see Fig. \ref{fig:FIGS31}) of the angle between the tangent at the colony boundary (green arrows) and the cell orientation (red arrows, with black dots representing the centroids) reveals that most cells tend to align tangentially along the colony boundary. The substrate stiffness however influences the extent to which this local alignment is achieved: cells growing on stiffer substrates exhibit a higher degree of tangential alignment, whereas low alignment is observed in case of cells growing on softer substrates, resulting in higher boundary roughness.

\subsubsection*{Nematic order parameter}
Cell orientation in phase-contrast images was computed using the structure tensor method, with local averaging performed using a Gaussian window approximately one-quarter the size of a single cell. Orientation angles were extracted via eigenvalue analysis of the tensor. The local nematic order parameter $S$ was calculated within square regions R (containing ~3–4 cells) using a moving grid approach. Only pixels within bacterial cells were considered to exclude boundary artifacts. The nematic order parameter for colonies growing on substrates with different Young's modulus is shown in Fig. \ref{fig:FIGSOP}. Details of the algorithm is provided in Ref. \cite{Rani2024}.

\subsection*{Statistical tests}
Statistical analysis was performed using two-sample t-tests across different substrate stiffness values, given by their Young's modulus. Data are presented as mean ± standard deviation (SD). Statistical significance was defined as $p < 0.05$. The asterisks on the plots indicate significance levels: $p < 0.05$ ($^*$) and $p < 0.01$ ($^{**}$). All analyses were conducted using Python (SciPy and NumPy).

\section*{Acknowledgments}
 
 We gratefully acknowledge the support from Human Frontier Science Program Cross-Disciplinary Fellowship (LT 00230/2021-C to G.R.) and the Institute for Advanced Studies, University of Luxembourg (AUDACITY Grant: IAS-20/CAMEOS to A.S.). We thank René Riedel for the fruitful discussions and supporting with the Young's modulus measurement on 0.75\% conc. agarose sample (additional Young's modulus data covered here have been reported elsewhere by the authors). A.S. thanks the Luxembourg National Research Fund for the ATTRACT Investigator Grant (A17/MS/11572821/MBRACE) for supporting this work.   

\bibliography{paper.bib}

\bibliographystyle{unsrt}

\end{document}